\documentclass[conference]{IEEEtran}
\usepackage[pdftex]{graphicx}
\usepackage{amsmath}
\usepackage{amssymb}
\graphicspath{{figures/}}

\begin{document}
\title{Frequency-Space Prediction Filtering for Phase Aberration Correction in Plane-Wave Ultrasound}

\author{\
\IEEEauthorblockN{Mostafa~Sharifzadeh, Habib~Benali, and Hassan~Rivaz}
\IEEEauthorblockA{Department of Electrical and Computer Engineering\\Concordia University\\ Montreal, QC, Canada\\mostafa.sharifzadeh@mail.concordia.ca}}
\maketitle

\begin{abstract}
Ultrasound imaging often suffers from image degradation stemming from phase aberration, which represents a significant contributing factor to the overall image degradation in ultrasound imaging. Frequency-space prediction filtering or FXPF is a technique that has been applied within focused ultrasound imaging to alleviate the phase aberration effect. It presupposes the existence of an autoregressive (AR) model across the signals received at the transducer elements and removes any components that do not conform to the established model.
In this study, we illustrate the challenge of applying this technique to plane-wave imaging, where, at shallower depths, signals from more distant elements lose relevance, and a fewer number of elements contribute to image reconstruction. While the number of contributing signals varies, adopting a fixed-order AR model across all depths, results in suboptimal performance. To address this challenge, we propose an AR model with an adaptive order and quantify its effectiveness using contrast and generalized contrast-to-noise ratio metrics.
\end{abstract}

\IEEEpeerreviewmaketitle

\section{Introduction}
Ultrasound stands as a commonly used modality in medical imaging owing to its numerous advantages, including portability, non-invasiveness, high temporal resolution, and affordability. Nonetheless, it often suffers from artifacts, with phase aberration as one of the main contributors to the degradation of image quality \cite{Pinton2011}.

The phase aberration effect originates from the spatially varying sound speed as sound waves travel through a heterogeneous medium. This effect introduces distortions in focused imaging by altering the focal point and perturbing the flat wavefront propagation in plane-wave imaging during transmission. Moreover, during reception, it hinders the coherent summation of echo signals in both focused and plane-wave imaging modes. Collectively, these factors lead to suboptimal image quality.

Numerous techniques have been proposed to compensate for the phase aberration effect in ultrasound images, all aimed at improving the performance of this modality, which could result in enhanced medical condition diagnosis, treatment, and image-guided interventions. These techniques include estimating delay errors by maximizing the cross-correlation between RF signals received at adjacent array elements \cite{Flax1988}, maximizing mean speckle brightness in a region of interest \cite{Nock1989}, using the generalized coherence factor for reducing focusing errors \cite{Li2003}, finding optimal sound speeds for subsequent imaging that maximizes the focus quality by analyzing lateral spatial frequency content in reconstructed images \cite{Napolitano2006}, assuming a spatially varying near-field phase screen and employing multistatic synthetic aperture data to perform the correction at each point adaptively \cite{Chau2019}, decoupling aberrations undergone by outgoing and incoming waves using the distortion matrix derived from the focused reflection matrix \cite{Lambert2021, Lambert2021a}, and incorporating singular value decomposition to design a beamformer robust to this artifact \cite{Bendjador2020}.
Another avenue of exploration involves leveraging the spatial distribution of sound speed within a given medium to mitigate the phase aberration effect \cite{Ali2022}. The spatial distribution can be estimated by exploiting phase shifts across a sequence of beamformed plane-wave images \cite{Stahli2020}, establishing a connection between the local sound speed along a wave propagation path and the average sound speed over that path \cite{Jakovljevic2018}, and estimating a global average sound speed to avoid spatial ambiguity issues \cite{Brevett2022}. Furthermore, convolutional neural networks have been employed to estimate the aberration profile \cite{Sharifzadeh2020} and to compensate for the phase aberration effect without reference non-aberrated data \cite{Sharifzadeh2023}.

Among these methods, the frequency-space (F-X) prediction filtering (FXPF), originally developed for random noise suppression in seismic imaging \cite{Canales1984}, has recently found application in correcting phase aberrations within focused ultrasound imaging \cite{Shin2018a}. This approach assumes an autoregressive (AR) model of order $p$ across the signals received by the transducer elements, systematically eliminating any components that deviate from the established model.
In this study, we demonstrate the challenge of applying this technique to plane-wave imaging, where adopting a fixed-order AR model across all depths results in suboptimal performance. Moreover, we propose the adaptive FXPF method to surmount this challenge. This method employs an AR model with an adaptive order according to the image depth.

\section{Methodology}
\subsection{Adaptive FXPF}
Let us consider a transducer with N elements and denote the Fourier transform of the received RF signal at time $t$ by element $n \in [1, N]$ located at $x_n$ as 
$RF_{n}(f)=\mathcal{F}\{RF(x_{n},t)\}$.
The FXPF establishes an AR model of order $p$ across the channel RF signals received at transducer elements. Specifically, in the frequency domain and for each temporal frequency $f_k$, the method predicts a signal as a linear combination of the signals received by the $p$ preceding channels:
\begin{equation}
	\label{eq-1}
	\begin{aligned}
		RF_{n+1}(f_k)=a_1(f_k)RF_{n}(f_k)+a_2(f_k)RF_{n-1}(f_k)+\\a_3(f_k)RF_{n-2}(f_k)+...+a_p(f_k)RF_{n+1-p}(f_k),
	\end{aligned}
\end{equation}
where coefficients denoted by $a$ need to be estimated. Given that Equation (\ref{eq-1}) represents a convolution, it can be expressed as
\begin{equation}
	\label{eq-2}
	RF_{n+1}
	= \begin{bmatrix}
		RF_{n} & RF_{n-1} & \dots & RF_{n+1-p}
	\end{bmatrix}
	\begin{bmatrix}
		a_1 \\ \vdots \\ a_p
	\end{bmatrix},
\end{equation}
where the $f_k$ was left out for simplicity of notation, but the equation pertains to a specific temporal frequency, denoted as $f_k$. Equation (\ref{eq-2}) can be written in a more general form as the product of a matrix and a vector. For example, when $p=4$, it can be written as
\begin{equation}
	\label{eq-3}
	\begin{bmatrix}
		RF_2 \\ RF_3 \\ RF_4 \\ RF_5 \\ \vdots \\ RF_{n+1} \\ 0 \\ 0 \\ 0
	\end{bmatrix}
	= \begin{bmatrix}
		RF_1 & 0 & 0 & 0 \\
		RF_2 & RF_1 & 0 & 0 \\
		RF_3 & RF_2 & RF_1 & 0 \\
		RF_4 & RF_3 & RF_2 & RF_1 \\
		\vdots & \vdots & \vdots & \vdots \\
		RF_n & RF_{n-1} & RF_{n-2} & RF_{n-3} \\
		0 & RF_{n} & RF_{n-1} & RF_{n-2} \\
		0 & 0 & RF_{n} & RF_{n-1} \\
		0 & 0 & 0 & RF_{n}
	\end{bmatrix}
	\begin{bmatrix}
		a_1 \\ a_2 \\ a_3 \\ a_4
	\end{bmatrix}.
\end{equation}
Let us express Equation (\ref{eq-3}) as
\begin{equation}
	\label{eq-4}
	\mathbf{d} = \mathbf{Ma},
\end{equation}
where $\mathbf{d}$ represents the vector comprising values associated with the current elements, $\mathbf{M}$ denotes the convolution matrix consisting of values corresponding to the preceding elements, and $\mathbf{a}$ is the prediction error filter with a length of $p$. In practice, channel RF data are inevitably contaminated with random noise from various sources. Therefore, the prediction error filter $\mathbf{a}$ in Equation (\ref{eq-4}) must be estimated from the noisy data $\mathbf{d}$. Achieving this requires minimizing the energy associated with the prediction error:
\begin{equation}
	\label{eq-5}
	\mathbf{L} = \|\mathbf{Ma-d}\|^2_2,
\end{equation}
where $\|.\|^2_2$ is the square of the Euclidean norm.
To minimize the cost function $\mathbf{L}$, it is required to set $\frac{\partial \mathbf{L}}{\partial \mathbf{a}} = 0$, which results in
\begin{equation}
	\label{eq-6}
	\mathbf{M^{T}d} = \mathbf{M^{T}Ma}.
\end{equation}
We can obtain an estimate $\mathbf{\hat{a}}$ of the prediction error filter $\mathbf{a}$ as
\begin{equation}
	\label{eq-7}
	\mathbf{\hat{a}} = (\mathbf{M^{T}M}+\mu\mathbf{I})^{-1}\mathbf{M^Td}
\end{equation}
where a stability factor $\mu$ is added into the diagonal components of $\mathbf{M^{T}M}$ to enhance the stability of the matrix inversion. In this study, $\mu$ was set to 0.01, and the results exhibit minimal sensitivity to its value.
After obtaining the estimated prediction error filter $\mathbf{\hat{a}}$, an estimate $\mathbf{\hat{d}}$ of the noise-free signal $\mathbf{d}$ can be acquired by applying it to the noisy data $\mathbf{M}$:
\begin{equation}
	\label{eq-8}
	\mathbf{\hat{d}} = \mathbf{M\hat{a}},
\end{equation}
where components of noisy data that do not conform to the established AR model are filtered out. Finally, the filtered RF signals can be obtained by applying the inverse Fourier transform.

While FXPF has been utilized for phase aberration correction in focused images \cite{Shin2018a}, employing this method for plane-wave images poses a challenge. This challenge primarily arises from the substantial variation in channel data across elements at shallower depths, where signals from more distant elements become irrelevant and may negatively impact the performance of the AR model. Even after applying apodization, using a high-order AR model for shallow depths with only a few echo signals may lead to over-smoothing during prediction filtering. In such scenarios, adopting a fixed-order AR model across all depths would result in suboptimal performance. To address this issue, we propose the utilization of an AR model with an adaptive order, defined as follows:

\begin{equation}
	\label{eq-9}
	p(z) = \min(p_{max},\ \lceil p_{max} \times \left(\frac{z}{f \times L}\right)^\beta \rceil),
\end{equation}
where $f$ represents the $f$-number, $z$ is the depth, $p_{max}$ is the maximum order used at depths where all elements are utilized for reconstruction, $\lceil . \rceil$ denotes rounding up to the nearest integer, and $\beta$ is the non-linearity coefficient that controls the speed of transition from lower orders to higher orders.
In summary, as per the formulation given by Equation (\ref{eq-9}), the AR model featuring an adaptive order always commences with a lower order (e.g., $p = 1$) for shallower depths, progressively increasing the order until it reaches $p_{max}$, a value we established for the deepest depths.

Although the technique was presented based on a forward AR model, a backward AR model can also be established by reversing the sequence of transducer elements \cite{Shin2018a}. To minimize potential directional biases and enhance the performance of the technique, the data underwent two independent filtering processes using both forward and backward AR models. The final output was then determined by averaging the results of these dual filtering paths. In practice, we used a moving axial kernel to compute the fast Fourier transform. Rather than processing the entire image all at once, we progressively shifted the kernel along the axial direction until the full depth was covered. Furthermore, once the method has been applied to the image for an initial iteration, it can undergo subsequent iterations, as long as it continues to yield improved outcomes. In our experimental setup, we set the $f$-number to $1.75$, employed an axial kernel size equivalent to one wavelength, and applied the FXPF method for 2 iterations. For the adaptive FXPF, we configured $p_{max}$ to be $4$, while $\beta$ was set to $1/3$. These specific parameters were selected due to their production of the most optimal results in our cases.

\subsection{Tissue-Mimicking Phantom Data}
An L11-5v linear array transducer was operated using a Vantage 256 system (Verasonics, Kirkland, WA) to acquire a single plane-wave image from a multi-purpose multi-tissue ultrasound phantom (Model 040GSE, CIRS, Norfolk, VA). The center and sampling frequencies were set at 5.208 MHz and 20.832 MHz, respectively, with the sound speed assumed to be 1540 m/s.
The transducer settings are summarized in Table~\ref{tbl-prob-settings}. We introduced a quasi-physical aberration \cite{Sharifzadeh2023} to the image by programming the scanner to excite transducer elements asynchronously according to a randomly generated aberration profile \cite{Sharifzadeh2020}. Moreover, delay errors introduced by the aberration profile were taken into account during the reception process for reconstructing the image. Received signals were stored as RF channel data after applying beamforming delays, serving as the input for the proposed method.
\begin{table}[h]
	\caption{The settings of linear array transducer L11-5v}
	\label{tbl-prob-settings}
	\setlength{\tabcolsep}{8pt}
	\def\arraystretch{1}%
	\begin{tabular}{p{115pt}p{50pt}p{40pt}}
		\hline
		\vspace{2pt}
		\textbf{Parameter}&
		\vspace{2pt}
		\textbf{Value}&
		\vspace{2pt}
		\textbf{Unit}
		\vspace{2pt}\\
		\hline
		\vspace{0pt}
		Number of Elements& 
		\vspace{0pt}
		128&
		\vspace{0pt}
		elements\\
		Elevation Focus& 
		20&
		mm\\
		Element Height& 
		5&
		mm\\
		Element Width& 
		0.27&
		mm\\
		Kerf& 
		0.03&
		mm\vspace{2pt}\\
		\hline
	\end{tabular}
\end{table}

\subsection{Quality Metrics}
\label{ssec:QualityMetrics}
To quantitatively measure the quality of reconstructed images, we calculated contrast and generalized contrast-to-noise ratio (gCNR) \cite{Rodriguez-Molares2020} metrics:
\begin{equation}
	Contrast=-20\log_\mathrm{10}(\frac{\mu_{t}}{\mu_{b}})\label{eq1},
\end{equation}
\begin{equation}
	\label{eq-gcnr}
	gCNR=1-\int_{-\infty}^{+\infty}{\min\limits_{\substack{x}}\{p_{t}(x),\ p_{b}(x)\}dx},
\end{equation}
where $t$ and $b$ represent the target and background regions, respectively, and $\mu$ stands for the mean value. In (\ref{eq-gcnr}), $x$ denotes the image value at any given pixel, and $p(x)$ is the probability density function of the values taken by pixels of a region.

\section{Results and Discussion}
\begin{figure*}
	\centering
	\includegraphics[width=0.9999\linewidth]{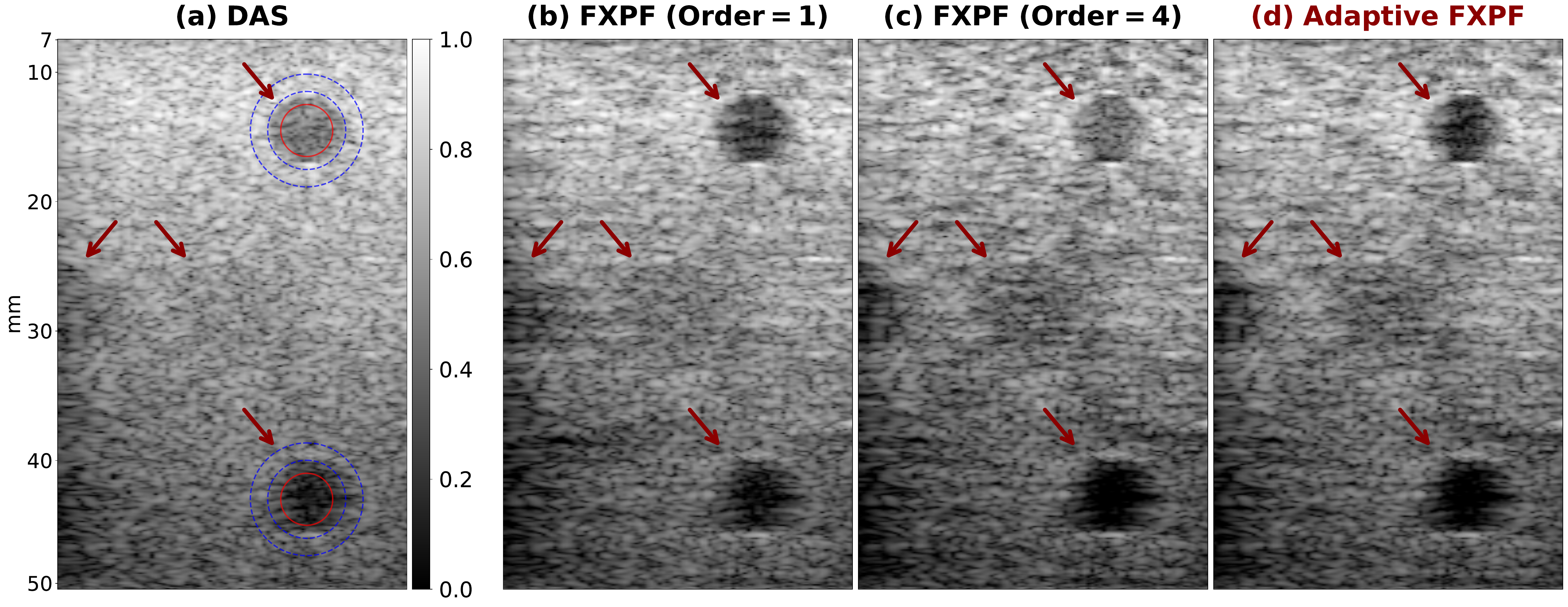}
	\caption{Qualitative comparison between FXPF methods with fixed orders and an adaptive order. (a) An aberrated single plane-wave image reconstructed using DAS. (b) The FXPF output with a fixed order of 1. (c) The FXPF output with a fixed order of 4. (d) The FXPF output with an adaptive order.}
	\label{fig1}
\end{figure*}

\begin{figure}
	\centering
	\includegraphics[width=0.9999\linewidth]{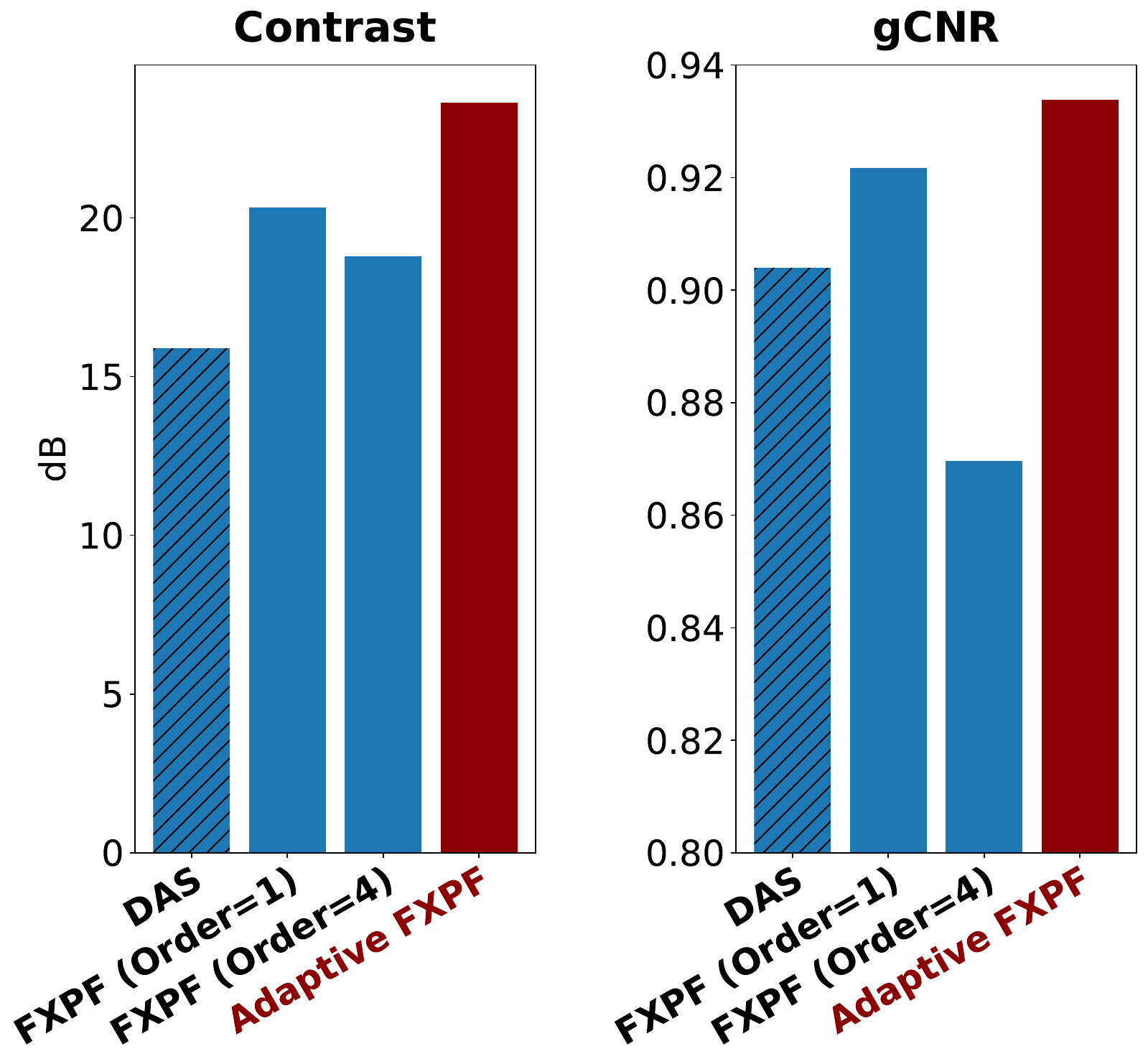}
	\caption{Quantitative comparison between FXPF methods with fixed orders and an adaptive order.}
	\label{fig2}
\end{figure}

Fig. \ref{fig1}(a) shows an aberrated single plane-wave image reconstructed using the delay-and-sum (DAS) method.
To mitigate the phase aberration effect, we applied the FXPF method with three distinct configurations. These include two AR models with fixed orders of 1 and 4, as well as an additional AR model incorporating the proposed adaptive order. The outputs obtained using fixed orders of 1 and 4 are illustrated in Fig. \ref{fig2}(b) and (c), respectively. While the model with a fixed order of 1 effectively enhanced the contrast of the anechoic cyst at shallow depths, it was nearly ineffective for the -6 dB and -3 dB hypoechoic cysts at the middle, as well as for the anechoic cyst at the bottom of the image. Conversely, the model with a fixed order of 4 improved the quality of the deeper cysts but degraded the contrast of the top cyst.
The output of the adaptive FXPF is shown in (d), highlighting a solution that effectively combines the advantages of both previous settings. This achievement was made possible by adaptively adjusting the order, utilizing a lower-order model for shallower depths, and progressively increasing the order for deeper depths.
Note that we generally observe more improvement for the bottom cyst when compared to the top one in all the images. This observation can be explained by the fact that the severity of the phase aberration effect, which requires correction, tends to be lower at shallower depths in contrast to deeper depths for two reasons. Firstly, perturbations in the wavefront become more pronounced as it propagates, resulting in an increased aberration effect during transmission as the wavefront advances. Secondly, as mentioned earlier, the aperture size is smaller at shallower depths, which mitigates the issue of incoherent summation at lower depths, as only a smaller number of neighboring elements are involved in the process of image reconstruction.
The contrast and gCNR metrics are calculated for the top and bottom anechoic cysts using the target and background regions shown in Fig. \ref{fig1}(a). Both metrics were calculated on the envelope-detected image in the linear domain before applying the log-compression, where the target region was inside the solid red circle and the background was the region between two dashed blue concentric circles. The average values across two cysts are reported in Fig. \ref{fig2}.

\section{Conclusion}
We demonstrated a challenge associated with phase aberration correction in plane-wave images using the FXPF method. This challenge arises due to substantial variations in channel data across elements at shallower depths, where signals from more distant elements lose relevance and can adversely affect the performance of the AR model.
To address this challenge, we proposed the adaptive FXPF, which adjusts the order of the AR model by employing a lower order for shallower depths and progressively increases the order for deeper depths. Both qualitative and quantitative results indicated that the adaptive approach provides higher performance in correcting the phase aberration effect.

\section*{Acknowledgment}
The authors would like to thank Natural Sciences and Engineering Research Council of Canada (NSERC) for funding.

\bibliographystyle{IEEEtran}
\bibliography{bibliography}

\end{document}